\documentclass[pdflatex,sn-mathphys-num]{sn-jnl}% Math and Physical Sciences Numbered Reference Style 
\usepackage{graphicx}%
\usepackage{multirow}%
\usepackage{amsmath,amssymb,amsfonts}%
\usepackage{amsthm}%
\usepackage{mathrsfs}%
\usepackage[title]{appendix}%
\usepackage{xcolor}%
\usepackage{textcomp}%
\usepackage{manyfoot}%
\usepackage{booktabs}%
\usepackage{algorithm}%
\usepackage{algorithmicx}%
\usepackage{algpseudocode}%
\usepackage{listings}%
\theoremstyle{thmstyleone}%
\theoremstyle{thmstyletwo}%

\theoremstyle{thmstylethree}%

\begin{document}

\title[Efficient Bitcoin Address Classification Using Quantum-Inspired Feature Selection]{Efficient Bitcoin Address Classification Using Quantum-Inspired Feature Selection}

\author*[1]{\fnm{Ming-Fong} \sur{Sie}}\email{d07922015@csie.ntu.edu.tw}
\author*[2,4]{\fnm{Yen-Jui} \sur{Chang}}\email{aceest@cycu.edu.tw}
\author[1]{\fnm{Chien-Lung} \sur{Lin}}\email{}
\author*[3,4]{\fnm{Ching-Ray} \sur{Chang}}\email{crchang@phys.ntu.edu.tw}
\author*[1]{\fnm{Shih-Wei} \sur{Liao}}\email{liao@csie.ntu.edu.tw}

\affil*[1]{\orgdiv{Department of Computer Science and Information Engineering}, \orgname{National Taiwan University}, \orgaddress{\street{No. 1, Sec. 4, Roosevelt Rd.}, \city{Taipei City}, \postcode{10617}, \state{Taiwan}}}
\affil[2]{\orgdiv{Master Program in Intelligent Computing and Big Data}, \orgname{Chung Yuan Christian University}, \orgaddress{\street{No. 499, Xingzhong N. Rd.}, \city{Taoyuan City}, \postcode{320317}, \state{Taiwan}}}
\affil[3]{\orgdiv{Department of Physics}, \orgname{National Taiwan University}} % 去掉了 orgaddress

\affil[4]{\orgdiv{Quantum Information Center}, \orgname{Chung Yuan Christian University}, \orgaddress{\street{No. 499, Xingzhong N. Rd.}, \city{Taoyuan City}, \postcode{320317}, \state{Taiwan}}}

\abstract{Over 900 million Bitcoin transactions have been recorded, posing considerable challenges for machine learning regarding computation time and maintaining prediction accuracy. We propose an innovative approach using quantum-inspired algorithms implemented with Simulated Annealing and Quantum Annealing to address the challenge of local minima in solution spaces. This method efficiently identifies key features linked to mixer addresses, significantly reducing model training time. By categorizing Bitcoin addresses into six classes—exchanges, faucets, gambling, marketplaces, mixers, and mining pools—and applying supervised learning methods, our results demonstrate that feature selection with SA reduced training time by 30.3\% compared to using all features in a random forest model while maintaining a 91\% F1-score for mixer addresses. This highlights the potential of quantum-inspired algorithms to swiftly and accurately identify high-risk Bitcoin addresses based on transaction features.}


\maketitle

\section{Introduction}

As traditional computing progresses, it is approaching the boundaries of its capabilities when faced with complex challenges like optimization or quantum phenomena simulation. To address these challenges, quantum-inspired computing has emerged, drawing inspiration from quantum mechanics to explore new dimensions of computational power. Quantum-inspired Ising machines offer a unique approach to solving integer programming problems by transforming them into equivalent Ising model representations. The Ising model \cite{brush1967history}, originally formulated to describe magnetic interactions, represents atoms (or spins) in two states (+1 or -1), with their interactions determining the system's overall energy. Many optimization problems can be translated into an energy minimization problem of the Ising model, with quantum or quantum-inspired methods used to find the configuration with the lowest energy.

Quantum-inspired algorithms leverage principles like superposition and tunneling to efficiently navigate the energy landscape of the Ising model. This enables them to find optimal solutions more effectively, especially in vast solution spaces with numerous local optima, bridging the gap between classical and quantum computing \cite{chang2024quantum, mandal2021approach}.

Bitcoin was launched in 2009, and it was a game-changer in the world of cryptocurrency\cite{nakamoto2008bitcoin}. It introduced a decentralized ledger that serves two main purposes: as a medium of exchange and a store of value\cite{baur2021volatility}. However, certain services known as "mixers" make Bitcoin transactions difficult to trace. Mixers are used to combine multiple users' cryptocurrency into target wallets, complicating the tracing process back to the original source\cite{rathore2022mixers}. Money launderers can use mixer services to hide illegal activities by converting their illicit proceeds into Bitcoin and using mixer services to obscure the origin\cite{liu2021detecting}.

Bitcoin's blockchain has recorded over 900 million transactions, making data analysis tasks such as predicting and understanding transaction behavior significantly challenging. Applying machine learning algorithms to this dataset presents hurdles like dataset imbalance\cite{ashfaq2022machine}, high dimensionality, and dynamic evolution. Dataset imbalance occurs when certain transaction types significantly outnumber others, leading to difficulty in obtaining accurate predictions. The complex features associated with each transaction lead to high dimensionality, and the continuous evolution of the data structure results in dynamic changes over time. Consequently, machine learning methods often incur high computational costs without guaranteed accuracy improvements. Given these challenges, exploring alternative approaches tailored to blockchain data is essential. Mixer addresses are crucial to regulatory bodies and entities aiming to maintain blockchain integrity\cite{wu2021towards}. Chang et al. \cite{chang2023prospects} provide a comprehensive overview of quantum computing applications in traditional and blockchain financial systems.

\setcounter{footnote}{1} % Set the footnote counter to start from 1
\begin{figure}[htbp]
\centerline{\includegraphics[width=12cm]{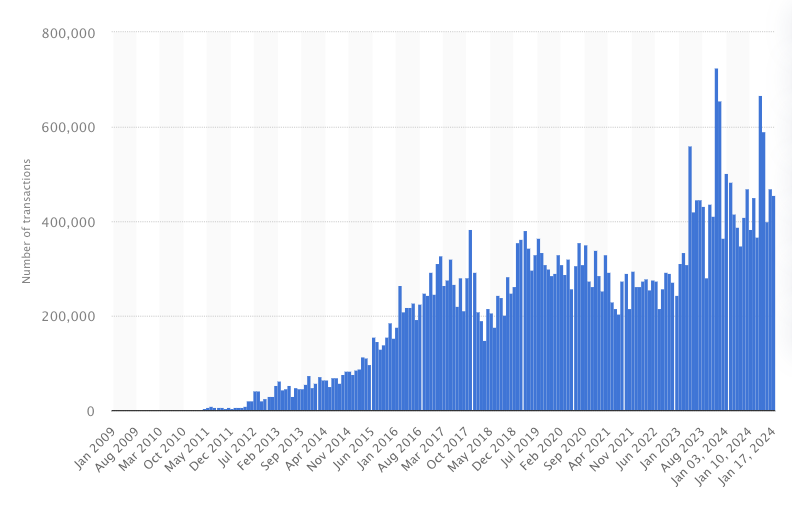}}
\caption{Number of daily transactions on the blockchain of Bitcoin from January 2009 to January 17, retrieved from \textit{Statista.com}.}
\label{fig:bitcoin_daily_transactions}
\end{figure}
\footnotetext{\url{https://www.statista.com/statistics/730806/daily-number-of-bitcoin-transactions/}}

We categorized Bitcoin addresses into various groups, such as exchanges, faucets, gambling, marketplaces, mixers, and mining pools, to create a structured framework for analysis. Our main focus is on mixer addresses, which are especially interesting because of their potential use in illegal activities and the privacy implications. Mixer addresses are used to obscure the trail of Bitcoin transactions, making them crucial to both regulatory bodies and entities that aim to maintain the blockchain's integrity.\cite{wu2021towards}

Our research proposes Quantum-Inspired Feature Selection (QIFS) that employs a Quadratic Unconstrained Binary Optimization (QUBO)-like model, aiming to balance feature influence and dependence for optimal classification performance to simplify the feature selection process for Bitcoin transaction records and use supervised learning techniques to construct a predictive model for the identification of mixer addresses, with a particular emphasis on random forests\cite{breiman2001random} and gradient-boosting decision trees\cite{chen2016xgboost}. These methods are complemented by cross-validation to ensure the reliability and generalizability of our model. The efficacy of our approach is evaluated using the receiver operating characteristic (ROC) curve\cite{zweig1993receiver}, a tool to assess the performance of classification models.

Our study has produced promising findings that demonstrate a significant reduction in the training time of the random forest model—by 30.3\%—while maintaining a 91\% F1-score in identifying mixer addresses. These results not only validate the effectiveness of our quantum-inspired algorithms with QUBO to select features to accelerate training time but also highlight its potential for practical applications, especially in identifying high-risk Bitcoin addresses. 

%Our research contributes to the literature on detecting Bitcoin fraud through address-based transaction history analysis. We propose and implement a solution framework that utilizes Simulated Annealing (SA)\cite{milne2017optimal}, Quantum Annealing (QA)\cite{finnila1994quantum}, and Quantum Annealing Binary Quadratic Model (QA BQM) to filter features of the on-chain transaction history of Bitcoin. This study demonstrates the potential of integrating quantum-inspired algorithms with machine learning, supporting the future development of scalable models.

\section{Method} 
\subsection{Quantum-Inspired Feature Selection and Optimization}

Large blockchain datasets often contain redundant information, leading to increased computational costs. Feature selection aims to distill essential information for efficient analysis. Our QIFS method identifies the most relevant features for Bitcoin mixer activity from transaction data, optimizing feature influence and independence through a QUBO model\cite{milne2017optimal}.

The cleaned blockchain transaction data is represented as a feature matrix \( \mathscr{D}_{m \times n} \), where each of the \( n \) columns corresponds to a feature, and each of the \( m \) rows represents transaction data associated with a Bitcoin address:

\[
\mathscr{D} = \begin{bmatrix}
 D_{11} & D_{12} & \cdots & D_{1n} \\
 D_{21} & D_{22} & \cdots & D_{2n} \\
 \vdots & \vdots & \ddots & \vdots \\
 D_{m1} & D_{m2} & \cdots & D_{mn}
\end{bmatrix}
\]

We also define an outcome vector \( \mathscr{O} \), which records the classification results for Bitcoin addresses:

\[
\mathscr{O} = \begin{bmatrix}
 O_1 \\
 O_2 \\
 \vdots \\
 O_m
\end{bmatrix}
\]

Here, \( O_{i} \) is a binary variable that takes on values 0 (rejection in Bitcoin classification) or 1 (acceptance).

We aim to select a subset of \( K \) features from the original \( n \) features to classify Bitcoin addresses. The QUBO model, adapted from Milne et al. \cite{milne2017optimal}, balances feature influence and dependence, using the Spearman rank correlation coefficient to evaluate the relevance and distinctiveness of each feature. 

The objective function is:

\[
  f(x) = -\alpha \sum_{j=1}^n x_j |\rho_{o_j}| + (1 - \alpha) \sum_{j=1}^n \sum_{\substack{k=1 \\ k \neq j}}^n x_j x_k |\rho_{jk}|,
\]

where \( x_j \) is a binary variable indicating if feature \( j \) is selected, and \( \alpha \) (where \( 0 \leq \alpha \leq 1 \)) weights the influence term. The negative sign in the first term aims to maximize influence, while the second term seeks to minimize dependence.

This cost function is optimized using a quantum-inspired machine to determine the feature subset that best balances the influence-dependence trade-off:

\[
  x^* = \arg \min_{\mathbf{X}} \left[ -\mathbf{X}^\top \mathbf{Q} \mathbf{X} \right],
\]

where \( \mathbf{Q} \) is a matrix encoding both influence and independence factors.

\begin{figure}[htbp]
\centerline{\includegraphics[width=15cm]{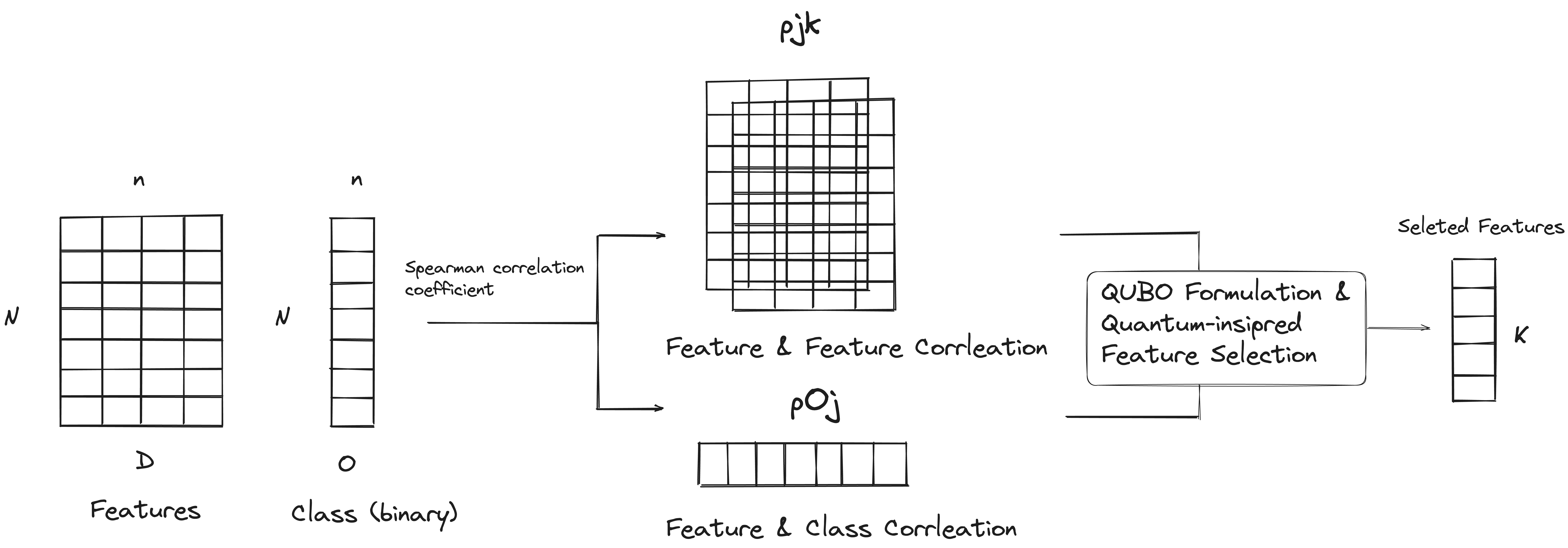}}
\caption{Quantum-Inspired Feature Selection pipeline.}
\label{fig:qifs_pipeline}
\end{figure}

Figure~\ref{fig:qifs_pipeline} illustrates our proposed Quantum-Inspired Feature Selection pipeline.

\subsection{Quantum-Inspired Acceleration and Machine Learning}

Quantum-inspired algorithms employ heuristic methods that approximate quantum behaviors, such as tunneling, to enhance optimization processes \cite{Hauke2019Perspectives}. These methods leverage an expanded search space, allowing for more efficient exploration compared to classical methods, and can potentially escape local minima to improve identification of the global optimum.

Simulated Annealing (SA) \cite{kirkpatrick1983optimization} is another technique to optimize the QUBO formulation during feature selection. SA mimics slowly cooling a material to reach its lowest energy state. By gradually reducing the temperature, SA probabilistically explores the solution space, reducing the likelihood of becoming trapped in local minima and increasing the probability of finding a global optimum. In our methodology, we employ simulated annealing to enhance feature selection, particularly when dealing with large and complex feature spaces.

Quantum annealing (QA) leverages quantum hardware to solve QUBO formulations directly. By formulating our feature selection problem as a QUBO model and utilizing quantum annealing for optimization, we benefit from faster convergence to optimal solutions compared to classical optimization methods.

Quantum-inspired machines, SA, and QA are all applied to further optimize QUBO to identify the global minimum energy cost during feature selection, thereby accelerating the supervised classification process. In our approach, we implement all three techniques to determine the most effective feature selection strategy for our specific task.

In this study, we evaluate the effectiveness of features selected using QIFS, implemented via SA and quantum methods such as D-Wave's QA and Binary Quadratic Model (BQM). The feature sets are validated on Bitcoin address classification tasks using 10-fold cross-validation, focusing on identifying mixer classes. Metrics such as precision and AUC are used to demonstrate the robustness and utility of features selected by these methods.

\begin{figure}[htbp]
\centerline{\includegraphics[width=15cm]{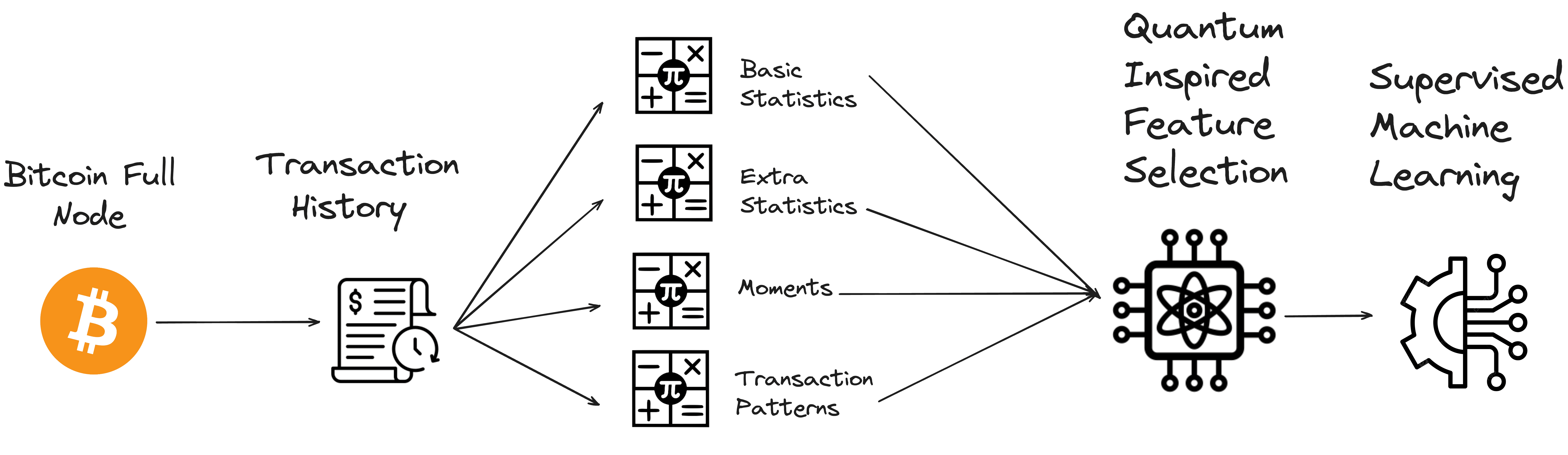}}
\caption{Quantum-Inspired Feature Selection overview.}
\label{fig:quantum_feature_selection}
\end{figure}

Figure~\ref{fig:quantum_feature_selection} provides an overview of our quantum-inspired approaches to feature selection, highlighting similarities and differences between the optimization techniques.

By leveraging quantum-inspired optimization, we aim to achieve high accuracy while minimizing computational overhead. These approaches significantly enhance processing efficiency, enabling effective classification even with large and complex datasets.

\section{Experiments} 
\begin{figure}[htbp]
\centerline{\includegraphics[width=10cm]{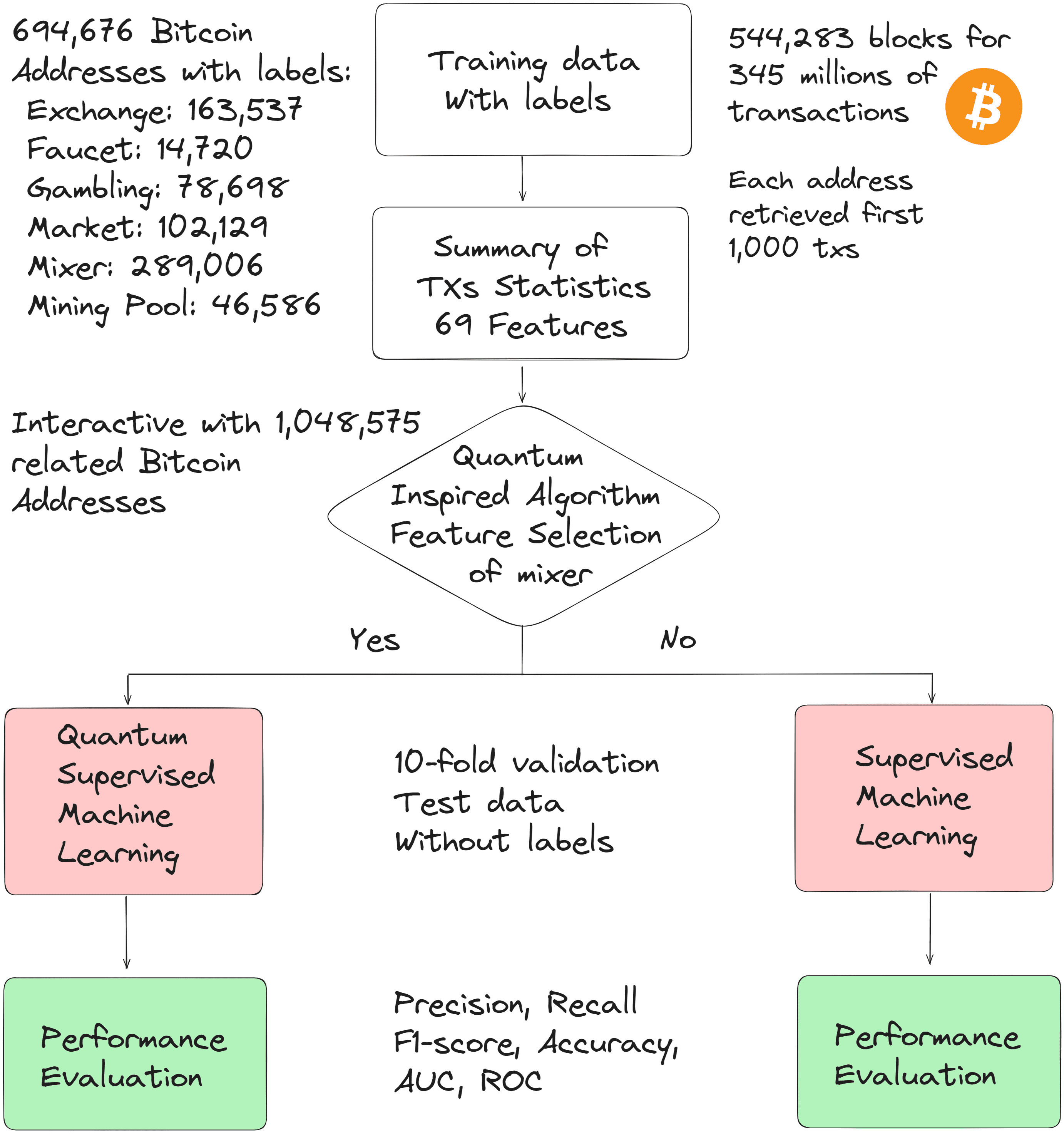}}
\caption{Experiment Flow}
\label{fig:Experiment Flow}
\end{figure}

\setcounter{footnote}{1} % Set the footnote counter to start from 1
Our experiment is divided into several distinct parts, as illustrated in Figure~\ref{fig:Experiment Flow} and implemented in code available at \href{https://github.com/Siemingfong/Quantom Annealing}{Siemingfong/Quantom Annealing}\footnote{\url{https://github.com/Siemingfong/Quantom_Annealing}}. 

% The first stage, detailed in Section~\ref{sec:data_collection}, focuses on collecting labeled data consisting of address-label pairs. Then it retrieves all transactions related to these addresses. The second stage, covered in Section~\ref{sec:smmarizing_transaction_history_Features}, addresses transaction history summarized into feature extraction. The third stage, described in Section~\ref{sec:quantum-inspired_implementation_details}, explores quantum-inspired SA, QA and QA BQM feature selection focusing on mixer class. The fourth stage involves quantum-inspired algorithm with supervised machine learning to train seven common classification algorithms, as detailed in Section~\ref{sec:training_classifiers} for the training classifier approach. Finally, the experiment concludes with performance evaluation, presented in Section~\ref{cap:evaluation_and_discussion}.

\subsection{Data Collection}
\label{sec:data_collection}

% First part is data collect. We setup Bitcoin fullnode to retrieved all the raw transaction history and we collect label Bitcoin addresses from WalletExploer building label Bitcoin addreesses in 6 classes include Exchange, Faucet, Gambling, Market, Mixer, Mining Pool catagories. 
\setcounter{footnote}{2} % Set the footnote counter to start from 1

We set up a full Bitcoin node leveraging 64GB of RAM, a 12th Gen Intel® Core™ i9-12900K CPU, GeForce GTX 1060 6 GB GPU, and 4 TB of SSD storage, granting us access to the complete raw transaction history of the network. This history contains detailed information about each transaction, including its hash, inputs, outputs, block height, and timestamp. The retrieval process took four months, covering 544,283 blocks and 345,882,038 transactions. Due to computational limitations, we focused on analyzing the first 1,000 transactions associated with each Bitcoin address. 

\subsection{Bitcoin Address Labeling}
We used \textit{WalletExplorer.com}\footnote{\href{https://walletexplorer.com/}{walletexplorer.com}} to acquire a dataset of 694,676 labeled Bitcoin addresses, interacting with 3,459,773 transactions. These addresses were categorized into six classes: exchange, faucet, gambling, market, mixer, and mining pool. Table \ref{table:dataset_details} summarizes the distribution of addresses and corresponding transactions across these categories.

\subsection{Smmarizing Transaction History Features}
\label{sec:smmarizing_transaction_history_Features}

We processed the transaction history for each address by aggregating all relevant transactions into a single sheet record. Then these records were sorted chronologically prior to feature extraction. Table~\ref{table:features} details the selected features, which are categorized into five groups: basic statistics features, extra statistics features, moment features and transaction patterns features. Calculating each feature requires access to the complete transaction history for an address. This requires retrieving all transactions involving the address before computing any statistical measures. By iterating through the sorted transaction history only once, we can efficiently extract all 69 features.

\begin{table}[t]
\caption{Dataset Details}
\centering
\begin{tabular}{lccc}
    \toprule
    Category & Addresses & Interactive Addresses & Interactive Transactions \\
    \midrule
    Exchange & 163,537 & 675,694 & 2,582,337 \\
    Faucet & 14,720 & 15,359 & 87,245 \\
    Gambling & 78,698 & 107,661 & 170,815 \\
    Market & 102,129 & 57,198 & 181,337 \\
    Mixer & 289,006 & 166,536 & 234,449 \\
    Pool & 46,586 & 26,127 & 203,590 \\
    \midrule
    Total & 694,676 & 1,048,575 & 3,459,773 \\
    \bottomrule
\end{tabular}
\label{table:dataset_details}
\vspace{-1em}
\end{table}

\begin{center}
\begin{table*}[!hb]
\caption{The List of Summarized Features from Transaction History.\label{table:features}}
\centering
\small
% \settowidth\tymin{$f_\mathrm{received}(10^i)$}
\begin{tabular*}{400pt}{@{\extracolsep\fill}ll@{\extracolsep\fill}}
  \toprule
  \textbf{Feature} & \textbf{Description} \\
  \midrule
  $f_\mathrm{TX}$ & Lifespan of Bitcoin address transaction frequency. \\
  $r_\mathrm{received}$ & Ratio of received transactions to total transactions. \\
  $r_\mathrm{coinbase}$ & Ratio of received coinbase transaction to total transactions. \\
  $f_\mathrm{spent}(10^i)$ & Frequency of digit $i$ transactions spent in USD, where $i$ ranges from $10^{-3}$ to $10^6$. \\
  $f_\mathrm{received}(10^i)$ & Frequency of digit $i$ transactions received in USD, where $i$ ranges from $10^{-3}$ to $10^6$. \\
  $r_\mathrm{payback}$ & Ratio of Bitcoin addresses in both inputs and outputs. \\
  $\bar{N}_\mathrm{inputs}$ & Average number of inputs used in spent transactions. \\
  $\bar{N}_\mathrm{outputs}$ & Average number of outputs used in spent transactions. \\
  \cmidrule(l{68pt}r{68pt}){1-2}
  \multicolumn{2}{@{}c@{}}{Basic Statistics} \\
  \midrule
  {$\textit{lifetime}$} & Lifespan of transactions days. \\
  {${\textit{BTC}_\mathrm{spent}}$} & Total Bitcoin amount spent. \\
  {${\textit{BTC}_\mathrm{received}}$} & Total Bitcoin amount received.\\
  {${\textit{USD}_\mathrm{spent}}$} & Total spent in USD, converted using daily BTC/USD rates from \textit{finance.yahoo.com}\footnote{\href{finance.yahoo.com/}{finance.yahoo.com/}}.\\
  \setcounter{footnote}{3}% Set the footnote counter to start from 3
  {${\textit{USD}_\mathrm{received}}$} & Total received in USD, converted using daily BTC/USD rates from \textit{finance.yahoo.com}\footnote{\href{finance.yahoo.com/}{finance.yahoo.com/}}. \\
  $n_\mathrm{TX}$ &  Total number of transactions. \\
  $n_\mathrm{spent}$ & Total number of spent transactions. \\
  $n_\mathrm{received}$ &  Total number of received transactions. \\
  $n_\mathrm{coinbase}$ & Total number of coinbase transactions. \\
  $n_\mathrm{payback}$ &  Total number of payback transactions. \\
  {$\mu_\mathrm{balance\_btc}$} & Average BTC amount held after each transaction. \\
  {$\sigma_\mathrm{balance\_btc}$} & Standard deviation of post-transaction BTC balances. \\
  {$\mu_\mathrm{balance\_usd}$} & Average USD amount held after each transaction. \\
  {$\sigma_\mathrm{balance\_usd}$} & Standard deviation of post-transaction USD balances. \\
  {$\sigma_{N_\mathrm{inputs}}$} & Standard deviation of total number of received transactions inputs. \\
  {$\sigma_{N_\mathrm{outputs}}$} & Standard deviation of total number of spent transactions outputs. \\
  \cmidrule(l{68pt}r{68pt}){1-2}
  \multicolumn{2}{@{}c@{}}{Extra Statistics} \\
  \midrule
  {$\mathit{m}_\mathrm{n,overall}$} & Distribution of the total transaction in moments. \\
  {$\mathit{m}_\mathrm{n,spent}$} & Distribution moments of spent transactions. \\
  {$\mathit{m}_\mathrm{n,received}$} & Distribution moments of received transactions. \\
  {$\mathit{m}_\mathrm{n,coinbase}$} & Distribution moments of coinbase transactions. \\
  {$\mathit{m}_\mathrm{n,payback}$} & Distribution moments of payback transactions. \\
  {$\mathit{m}_\mathrm{n,interval}$} & Distribution moments of transaction intervals. \\
  \cmidrule(l{68pt}r{68pt}){1-2}
  \multicolumn{2}{@{}c@{}}{Moments} \\
  \midrule
  {$\mathit{t}_\mathrm{n,input}$} & Total number of multiple input transactions.\\
  {$\mathit{t}_\mathrm{n,output}$} & Total number of multiple output transactions.\\
  {$\mathit{t}_\mathrm{overall}$} & Both transactions involve multiple inputs and outputs.\\
  \cmidrule(l{68pt}r{68pt}){1-2}
  \multicolumn{2}{@{}c@{}}{Tranascion Patterns} \\
  \bottomrule
\end{tabular*}
\end{table*}
\end{center}
\clearpage

The four transaction types, coinbase, spent, received, and payback, are mutually exclusive. A transaction is categorized as a coinbase transaction if it has a coinbase input, representing a block reward. If it lacks a coinbase input and the sender's address appears in any of its inputs, it's identified as a spent transaction. Conversely, if the address appears only in the outputs, it's classified as a received transaction. Finally, transactions with the address in both inputs and outputs are categorized as payback transactions. To understand the behavior of the address, we analyze the distribution moments of various types of transactions. These include total transactions, spent transactions, received transactions, coinbase transactions, payback transactions, and multiple input and output transactions. We believe that the frequency of transactions for each category helps identify address types. Beyond frequency, moments of these distributions and transaction intervals reveal further insight. For example, they can indicate an address that was initially active but has become dormant, or even exhibit periodic transaction patterns. In addition, distinguishing between spending, receiving, and payback transactions is crucial. Provides valuable clues on how addresses are being used within the Bitcoin transaction history.

\subsection{Quantum-Inspired Implementation}
\label{sec:quantum-inspired_implementation_details}

\begin{figure}[htbp]
\centerline{\includegraphics[width=12cm]{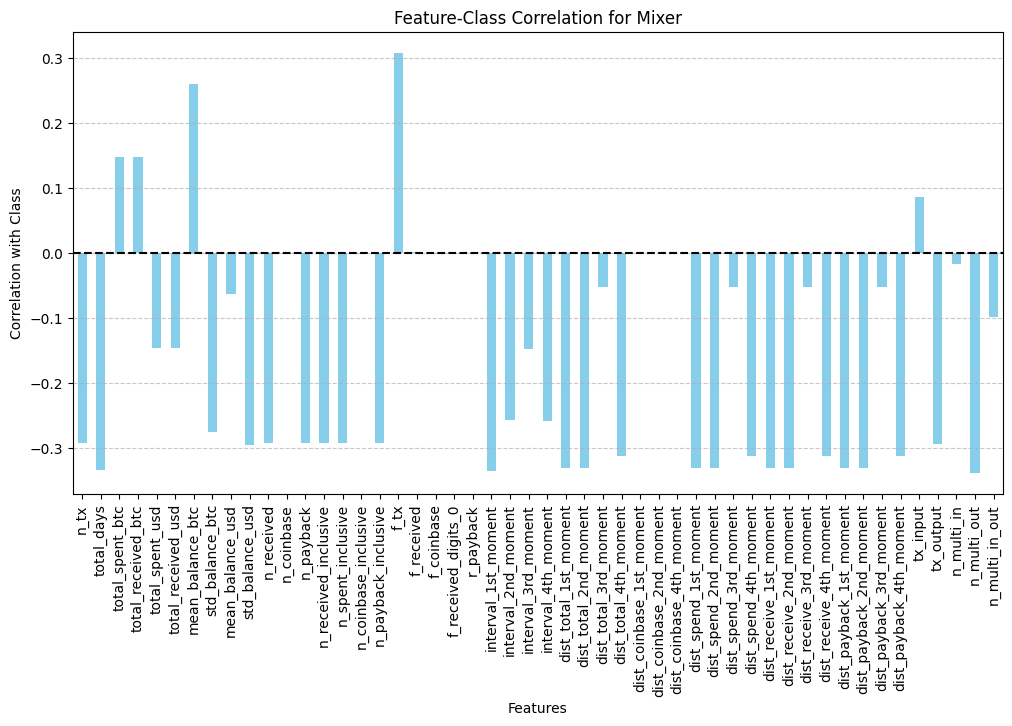}}
\caption{Mixer Spearman Correlations}
\label{fig:Mixer Spearman Correlations}
\end{figure}

Before the QISF process, each class is binarized as 0 or 1, transforming the data into a QUBO structure suitable for decision-making. Subsequently, the Spearman method is employed to compute correlations between each feature and the mixer class. The results of this correlation analysis are visualized in Figure~\ref{fig:Mixer Spearman Correlations}. Table~\ref{table:Quantum_Inspired_feature_select_time} evaluates the time efficiency of various QISF methods for the mixer class. SA, which selected 23 features, exhibited the lowest computational time at 0.0055 seconds. In comparison, QA with 9 features and QA BQM with 7 features required considerably more time, measuring 6.644 seconds and 6.0041 seconds, respectively. These results highlight that SA offers a significant computational speed advantage over other quantum-inspired algorithms in feature selection tasks.

\begin{table}[t]
\caption{\textbf{Quantum-Inspired feature selection time consumption in mixer class.}\label{table:Quantum_Inspired_feature_select_time}}
\centering
\begin{tabular}{lrr}
  \toprule
  \textbf{Quantum-Inspired Algorithms} & \textbf{Features} & \textbf{Time} \\
  \midrule
  Simulated Annealing & 23 & \textbf{0.0055 s} \\
  Quantum Annealing & 9 & 6.644 s \\
  Quantum Annealing BQM & 7 & 6.0041 s \\
  \bottomrule
\end{tabular}
\end{table}

The QISF methodology is specifically tailored for the mixer class. For implementation, we utilized the Python library dwave-neal \cite{dwave_neal}, which progressively converges toward the equilibrium distribution of the Ising model by iteratively updating spins in a sequence of increasing beta values. For quantum processing, the D-Wave Quantum Machine Advantage system 6.4, equipped with 5,612 working qubits, was employed. The D-Wave Hybrid solver (version 2.2) was used to solve general BQM problems. Tables~\ref{table:QUBO_23_features}, ~\ref{table:QA_7_features} provide details on the features selected by SA, QA, and QA BQM for the mixer class.

\begin{table}[t]
\caption{SA feature select in mixer class.\label{table:QUBO_23_features}}
\centering
\begin{minipage}{0.45\linewidth}
\centering
\textbf{SA top 10 features.}\\[1ex] % Add a little space after the text
\begin{tabular}{lr}
  \toprule
  \textbf{Feature Name} & \textbf{Feature Type} \\
  \midrule
  ${n_\mathrm{TX}}$ & Extra Stats \\
  ${n_\mathrm{payback}}$ & Extra Stats \\
  $n_\mathrm{received}$ & Extra Stats \\
  $n_\mathrm{spent}$ & Extra Stats \\
  $\textit{lifetime}$ & Extra Stats \\
  $f_\mathrm{received}(10^0)$ & Basic Stats \\
  $r_\mathrm{payback}$ & Basic Stats \\
  $\mathit{m}_\mathrm{2,interval}$ & Moments \\
  $\mathit{m}_\mathrm{3,interval}$ & Moments \\
  $\mathit{m}_\mathrm{2,overall}$ & Moments \\
  \bottomrule
\end{tabular}
\end{minipage}
% \hfill
\begin{minipage}{0.45\linewidth}
\centering
\textbf{SA top 11 to 23 features.}\\[1ex]
\begin{tabular}{lr}
  \toprule
  \textbf{Feature Name} & \textbf{Feature Type} \\
  \midrule
  ${\textit{BTC}_\mathrm{spent}}$ & Extra Stats \\
  $\mathit{m}_\mathrm{2,coinbase}$ & Moments \\
  $\mathit{m}_\mathrm{1,spent}$ & Moments \\
  $\mathit{m}_\mathrm{4,spent}$ & Moments \\
  $\mathit{m}_\mathrm{3,received}$ & Moments \\
  $\mathit{m}_\mathrm{4,received}$ & Moments \\
  $\mathit{m}_\mathrm{1,payback}$ & Moments \\
  $\mathit{m}_\mathrm{2,payback}$ & Moments \\
  $\mathit{m}_\mathrm{4,payback}$ & Moments \\
  ${\textit{USD}_\mathrm{spent}}$ & Extra Stats \\
  ${\textit{USD}_\mathrm{received}}$ & Extra Stats \\
  $\sigma_\mathrm{balance\_btc}$ & Extra Stats \\
  $\mu_\mathrm{balance\_usd}$ & Extra Stats \\
  \bottomrule
\end{tabular}
\end{minipage}
\end{table}

\begin{table}[t]
\caption{QA and QA BQM feature select in mixer class.\label{table:QA_7_features}}
\centering
\makebox[\textwidth][c]{%
\begin{minipage}{0.45\linewidth}
\centering
\textbf{QA top 9 features.}\\[1ex]
\begin{tabular}{lr}
  \toprule
  \textbf{Feature Name} & \textbf{Feature Type} \\
  \midrule
  ${\textit{USD}_\mathrm{received}}$ & Extra Stats \\
  ${n_\mathrm{coinbase}}$ & Extra Stats \\
  $f_\mathrm{TX}$ & Basic Stats \\
  $r_\mathrm{received}$ & Basic Stats \\
  $r_\mathrm{coinbase}$ & Basic Stats \\
  $\mathit{m}_\mathrm{2,interval}$ & Moments \\
  $\mathit{m}_\mathrm{4,received}$ & Moments \\
  $\mathit{t}_\mathrm{n,output}$ & TX Pattern \\
  $\mathit{t}_\mathrm{overall}$ & TX Pattern \\
  \bottomrule
\end{tabular}
\end{minipage}
\hfill
\begin{minipage}{0.45\linewidth}
\centering
\textbf{QA BQM top 7 features.}\\[1ex]
\begin{tabular}{lr}
  \toprule
  \textbf{Feature Name} & \textbf{Feature Type} \\
  \midrule
  $\textit{lifetime}$ & Extra Stats \\
  $\mu_\mathrm{balance\_btc}$ & Extra Stats \\
  ${n_\mathrm{coinbase}}$ & Extra Stats \\
  $f_\mathrm{TX}$ & Basic Stats \\
  $r_\mathrm{received}$ & Basic Stats \\
  $\mathit{m}_\mathrm{3,payback}$ & Moments \\
  $\mathit{t}_\mathrm{overall}$ & TX Pattern \\
  \bottomrule
\end{tabular}
\end{minipage}%
}
\end{table}

% \begin{table}[t]
% \caption{QA feature select in mixer class.\label{table:QA_7_features}}
% \centering
% \begin{minipage}{0.45\linewidth}
% \centering
% \textbf{The top 9 features.}\\[1ex] % Add a little space after the text
% \begin{tabular}{lr}
%   \toprule
%   \textbf{Feature Name} & \textbf{Feature Type} \\
%   \midrule
%   ${\textit{USD}_\mathrm{received}}$ & Extra Stats \\
%   ${n_\mathrm{coinbase}}$ & Extra Stats \\
%   $f_\mathrm{TX}$ & Basic Stats \\
%   $r_\mathrm{received}$ & Basic Stats \\
%   $r_\mathrm{coinbase}$ & Basic Stats \\
%   $\mathit{m}_\mathrm{2,interval}$ & Moments \\
%   $\mathit{m}_\mathrm{4,received}$ & Moments \\
%   $\mathit{t}_\mathrm{n,output}$ & TX Pattern \\
%   $\mathit{t}_\mathrm{overall}$ & TX Pattern \\
%   \bottomrule
% \end{tabular}
% \end{minipage}
% \end{table}

% \begin{table}[t]
% \caption{QA BQM feature select in mixer class.\label{table:QA_BQM_7_features}}
% \centering
% \begin{minipage}{0.45\linewidth}
% \centering
% \textbf{The top 7 features.}\\[1ex] % Add a little space after the text
% \begin{tabular}{lr}
%   \toprule
%   \textbf{Feature Name} & \textbf{Feature Type} \\
%   \midrule
%   $\textit{lifetime}$ & Extra Stats \\
%   $\mu_\mathrm{balance\_btc}$ & Extra Stats \\
%   ${n_\mathrm{coinbase}}$ & Extra Stats \\
%   $f_\mathrm{TX}$ & Basic Stats \\
%   $r_\mathrm{received}$ & Basic Stats \\
%   $\mathit{m}_\mathrm{3,payback}$ & Moments \\
%   $\mathit{t}_\mathrm{overall}$ & TX Pattern \\
%   \bottomrule
% \end{tabular}
% \end{minipage}
% \end{table}

\subsection{Training Classifiers}
\label{sec:training_classifiers}
We training of seven common classification algorithms include Logistic Regression\cite{lavalley2008logistic}, Adaptive Boosting with Decision Tree (AdaBoost-SAMME)\cite{freund1997decision, hastie2009multi}, Random Forest\cite{breiman2001random}, Extreme Gradient Boosting (XGBoost)\cite{chen2016xgboost}, LightGBM\cite{ke2017lightgbm}, SVM\cite{cortes1995support} and Neural Network. These algorithms serve as a baseline for comparison with the quantum approach. We leverage the Python machine learning library Scikit-learn\cite{pedregosa2011scikit} to employ seven different classifiers. To optimize each classifier and identify a suitable set of parameters, we utilize a 10-fold cross-validation. Notably, decision tree-based methods are unaffected by data normalization. Therefore, for these methods, we retain the original data. However, for classifiers like logistic regression and SVM, which rely on distance metrics, we normalize the features by dividing each dimension by its maximum absolute value.

Among the mixer classes, both Random Forest and SA Random Forest achieve the best F1 score, reaching 92\% and 91\%. SA Random Forest exhibits a faster 30.3\% training time 212 minutes and 23.2 seconds compared to Quantum Random Forest 305 minutes and 5.7 seconds. Table~\ref{table:result_mixer_supervised_and_quantum} details the results of both all features and SA feature selection of mixer in supervised classifiers. Our experiments revealed that most classifiers performed consistently on the mixer category. However, Logistic Regression showed poor performance on this category. This suggests that tree-based classifiers might be better suited for this task compared to linear models. 

\begin{table*}[t]
\caption{Supervised Classifiers with Full Features, Quantum-inspired Feature Selection in Mixer Class.\label{table:result_mixer_supervised_and_quantum}}
\footnotesize % 使用更小的字體
\setcounter{footnote}{0} % 重置註腳編號
\begin{center}
\resizebox{\textwidth}{!}{ % 縮放整個表格以適應頁面
\begin{tabular}{@{\extracolsep{\fill}}p{4cm}p{1.5cm}p{1.1cm}p{0.7cm}p{1.6cm}p{1cm}p{3cm}@{\extracolsep{\fill}}}
  \toprule
  \textbf{Method} & \textbf{Precision} $\uparrow$ & \textbf{Recall} $\uparrow$ & \textbf{F1} $\uparrow$ & \textbf{Accuracy} $\uparrow$ & \textbf{AUC} $\uparrow$ & \textbf{Training Time} $\downarrow$ \\
  \midrule
  Logistic Regression & 0.6 & 0.57 & 0.58 & 0.84 & 0.92 & 10 mins 56.8 s \\
  AdaBoost      & 0.4 & 0.81 & 0.53 & 0.84 & 0.74 & 1106 mins 52.8 s \\
  Random Forest       & \textbf{0.87} & 0.97 & \textbf{0.92} & \textbf{0.95} & \textbf{0.99} & 305 mins 5.7 s \\
  XGBoost             & 0.86 & 0.89 & 0.88 & \textbf{0.95} & \textbf{0.99} & 70 mins 5 s \\
  LightGBM            & 0.85 & 0.88 & 0.87 & 0.94 & 0.98 & 56 mins 1.8 s \\
  SVM *10\%\footnote{Using only 10\% of the dataset}            & 0.73 & 0.37 & 0.49 & 0.85 & 0.96 & 532 mins 58.9 s \\
  Neural Network *10\%\footnote{Neural Network trained with 10\% of the available data}            & 0.68 & 0.93 & 0.79 & 0.85 & 0.98 & 87 mins 51.5 s \\
  SA Logistic Regression & 0.53 & 0.8 & 0.59 & 0.91 & 0.83 & 10 mins 49.6 s \\
  SA AdaBoost      & 0.33 & 0.65 & 0.44 & 0.81 & 0.69 & 589 mins 21.1 s \\
  SA Random Forest       & \textcolor{red}{0.85} & \textcolor{red}{0.97} & \textcolor{red}{0.91} & \textcolor{red}{0.94} & \textcolor{red}{\textbf{0.99}} & \textcolor{red}{212 mins 23.2 s} \\
  SA XGBoost             & 0.54 & 0.63 & 0.58 & 0.87 & 0.89 & 67 mins 13.2 s \\
  SA LightGBM            & 0.81 & 0.84 & 0.83 & 0.94 & 0.96 & 37 mins 43.9 s \\
  SA SVM *10\%\footnote{SVM trained with 10\% of the dataset using SA}            & 0.22 & \textbf{1} & 0.36 & 0.42 & 0.71 & 4465 mins 27.2 s \\
  SA Neural Network *10\%\footnote{Neural Network trained with 10\% of the data using SA}            & 0.66 & 0.94 & 0.77 & 0.83 & 0.98 & 83 mins 3.7 s \\  
  QA Logistic Regression & 0.49 & 0.54 & 0.50 & 0.82 & 0.82 & 6 mins 47.8 s \\
  QA AdaBoost      & 0.57 & 0.49 & 0.52 & 0.85 & 0.84 & 513 mins 5.6 s \\
  QA Random Forest       & 0.58 & 0.62 & 0.60 & 0.86 & 0.88 & 125 mins 27.7 s \\
  QA XGBoost             & 0.54 & 0.63 & 0.58 & 0.87 & 0.89 & 63 mins 29.5 s \\
  QA LightGBM            & 0.55 & 0.62 & 0.58 & 0.87 & 0.89 & 37 mins 43.9 s \\
  QA SVM *10\%\footnote{SVM trained with 10\% of the dataset using QA}            & 0.26 & \textbf{1} & 0.41 & 0.51 & 0.86 & 4492 mins 26.3 s \\
  QA Neural Network *10\%\footnote{Neural Network trained with 10\% of the data using QA}            & 0.54 & 0.60 & 0.57 & 0.85 & 0.88 & 87 mins 7.3 s \\  
  QA BQM Logistic Regression & 0.51 & 0.56 & 0.53 & 0.85 & 0.84 & \textbf{6 mins 44.1 s} \\
  QA BQM AdaBoost      & 0.47 & 0.74 & 0.57 & 0.84 & 0.89 & 563 mins 56.8 s \\
  QA BQM Random Forest       & 0.65 & 0.76 & 0.75 & 0.90 & 0.93 & 183 mins 17.5 s \\
  QA BQM XGBoost             & 0.48 & 0.76 & 0.59 & 0.85 & 0.90 & 63 mins 36.6 s \\
  QA BQM LightGBM            & 0.48 & 0.75 & 0.58 & 0.85 & 0.90 & 38 mins 25 s \\
  QA BQM SVM *10\%\footnote{SVM trained with 10\% of the dataset using QA}            & 0.54 & 0.09 & 0.16 & 0.83 & 0.86 & 4391 mins 7.1 s \\
  QA BQM Neural Network *10\%\footnote{Neural Network trained with 10\% of the data using BQM}            & 0.54 & 0.58 & 0.56 & 0.85 & 0.88 & 83 mins 49.5 s \\ 
  \bottomrule
\end{tabular}
}
\end{center}
\end{table*}

\section{Performance Evaluation} \label{sec:performance_evaluation}

\begin{figure}[htbp]
    \centering
    \begin{minipage}{0.45\textwidth}
        \centering
        \includegraphics[width=\textwidth]{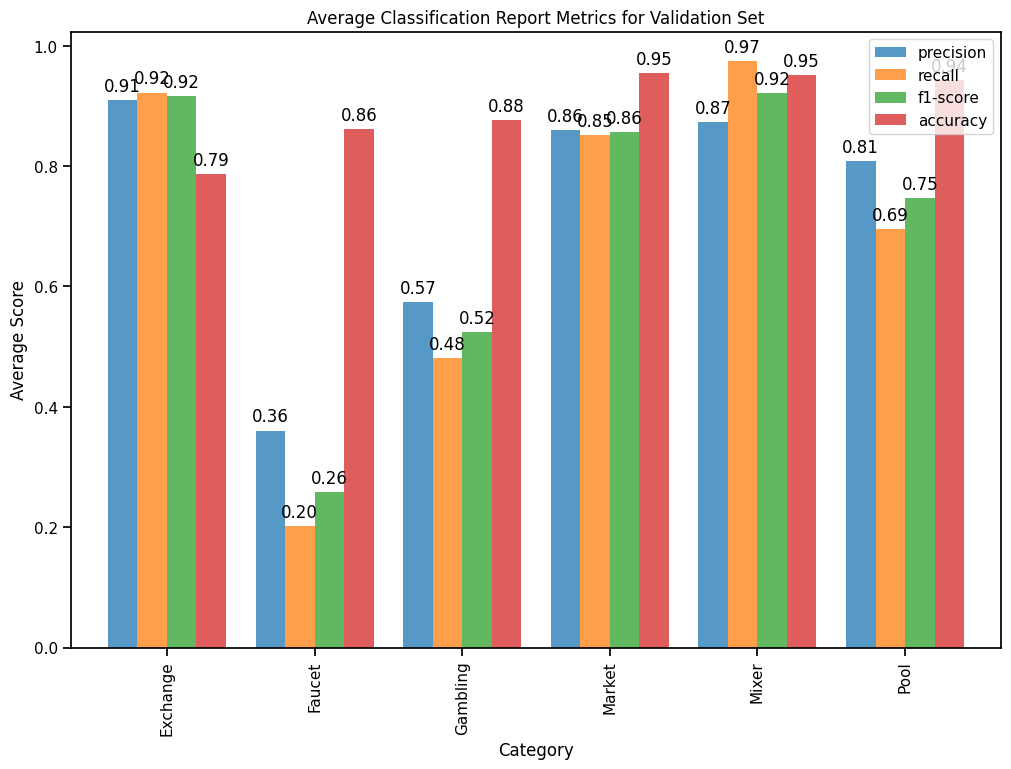}
        \caption{Random Forest with Each Class Evaluation}
        \label{fig:Random Forest with each class evaluation}
    \end{minipage}
    \hfill
    \begin{minipage}{0.45\textwidth}
        \centering
        \includegraphics[width=\textwidth]{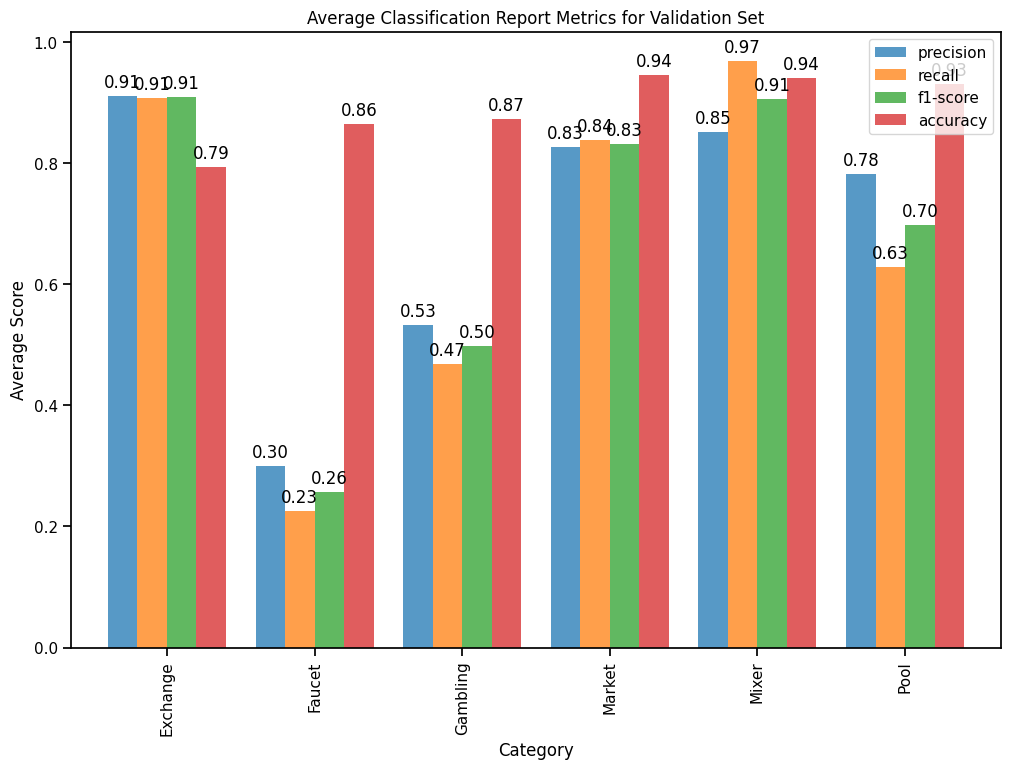}
        \caption{SA Random Forest with Each Class Evaluation}
        \label{fig:Simulate Annealing Random Forest with each class evaluation}
    \end{minipage}
\end{figure}

We analyze the precision, recall, and F1-score for the following 6 classes: Exchange, Faucet, Gambling, Market, Mixer, and Pool. Specifically for the Mixer class, both the Random Forest and the SA Random Forest models achieved the highest F1-score of 0.92 and 0.91 shown in Figure~\ref{fig:Random Forest with each class evaluation} and Figure~\ref{fig:Simulate Annealing Random Forest with each class evaluation}, indicating a strong and consistent classification performance. The QA Random Forest had a lower F1-score of 0.60, while the QA BQM Random Forest showed some improvement with an F1-score of 0.70.

Table~\ref{table:result_mixer_supervised_and_quantum} illustrates the performance of various supervised classifiers using full features, SA, QA and QA BQM for feature selection in the Mixer class. The classifiers evaluated include traditional models such as Logistic Regression and SVM, as well as ensemble methods like Random Forest, AdaBoost, and Gradient Boosting algorithms such as XGBoost and LightGBM. In several cases, models were trained on 10\% of the available dataset to compare their performance when data is limited. For instance, the Neural Network *10\% was trained using 10\% of the data, while models like SA SVM *10\% and QA SVM *10\% were trained with reduced datasets using SA and QA feature selection techniques, respectively. Each model’s precision, recall, F1 score, accuracy, AUC, and training time were recorded, highlighting the trade-offs between computational cost and performance.

Overall, Table~\ref{table:result_mixer_supervised_and_quantum} demonstrates that traditional machine learning models Random Forest, XGBoost, and LightGBM achieved the highest accuracy around 95\% and AUC around 0.99 within the mixer class. Notably, Random Forest exhibited the strongest F1-score 92\% and recall 0.97\% among all models. In terms of training time, LightGBM emerged as the fastest, taking only about 56 minutes, while AdaBoost was significantly slower, requiring over 1100 minutes. Interestingly, some SA for feature selection in Logistic Regression and Random Forest, achieved recall scores comparable to their classical counterparts. However, these quantum models generally exhibited lower precision and accuracy. Encouragingly, SA Random Forest showed a significant improvement in training time efficiency, offering a roughly 30\% reduction compared to the classical Random Forest while maintaining high F1-score and recall.

In different classes of bitcoin addresses, the Random Forest model, as demonstrated in Figure~\ref{fig:Random Forest with each class evaluation}, exhibited strong performance in certain categories. Specifically, it achieved excellent results for the Exchange class, with a precision of 0.91\%, recall of 0.92\%, and an F1-score of 0.92\%. Likewise, the model performed well in the Market class, reaching a precision of 0.86\%, recall of 0.85\%, and an F1-score of 0.85\%. For the Mixer class, the model showed particularly high recall 0.97\% and also maintained strong precision 0.87\% and an F1-score of 0.91\%. However, the model faced challenges with the Faucet class, where it struggled to achieve a precision of 0.36\%, recall of 0.20\%, and an F1-score of 0.26\%. In the Gambling class, the Random Forest showed a trade-off between precision 0.57\% and recall 0.48\%, leading to an F1-score of 0. 52\%. For the Pool class, the model's performance was more balanced, with a precision of 0.81\%, recall of 0.69\%, and an F1-score of 0.75\%, indicating moderate success across these metrics.

% \begin{figure}[htbp]
% \centerline{\includegraphics[width=15cm]{figures/rm_testing_recall_without_avg.png}}
% \caption{Random Forest with Each Class Evaluation}
% \label{fig:Random Forest with each class evaluation}
% \end{figure}

The SA Random Forest model, as shown in Figure~\ref{fig:Simulate Annealing Random Forest with each class evaluation}, demonstrated a performance similar to the Random Forest in the exchange, market and mixer classes, maintaining strong precision, recall and F1 scores. However, the SA Random Forest exhibited lower precision 0.30\% and recall 0.23\% for the Faucet class, further highlighting its challenges in this category compared to the standard model. In the Gambling class, the SA Random Forest also displayed a similar trade-off between precision 0.53\% and recall 0.47\%, leading to a marginally lower F1-score 0.50\%. Additionally, the SA Random Forest demonstrated lower recall 0.63\% for the Pool class compared to the Random Forest, although its precision remained relatively close at 0.78\%, resulting in an overall F1-score of 0.70\%.

% \begin{figure}[htbp]
% \centerline{\includegraphics[width=15cm]{figures/qrm_testing_recall_without_avg.png}}
% \caption{SA Random Forest with Each Class Evaluation}
% \label{fig:Simulate Annealing Random Forest with each class evaluation}
% \end{figure}

\begin{figure}[htbp]
    \centering
    \begin{minipage}{0.45\textwidth}
        \centering
        \includegraphics[width=\textwidth]{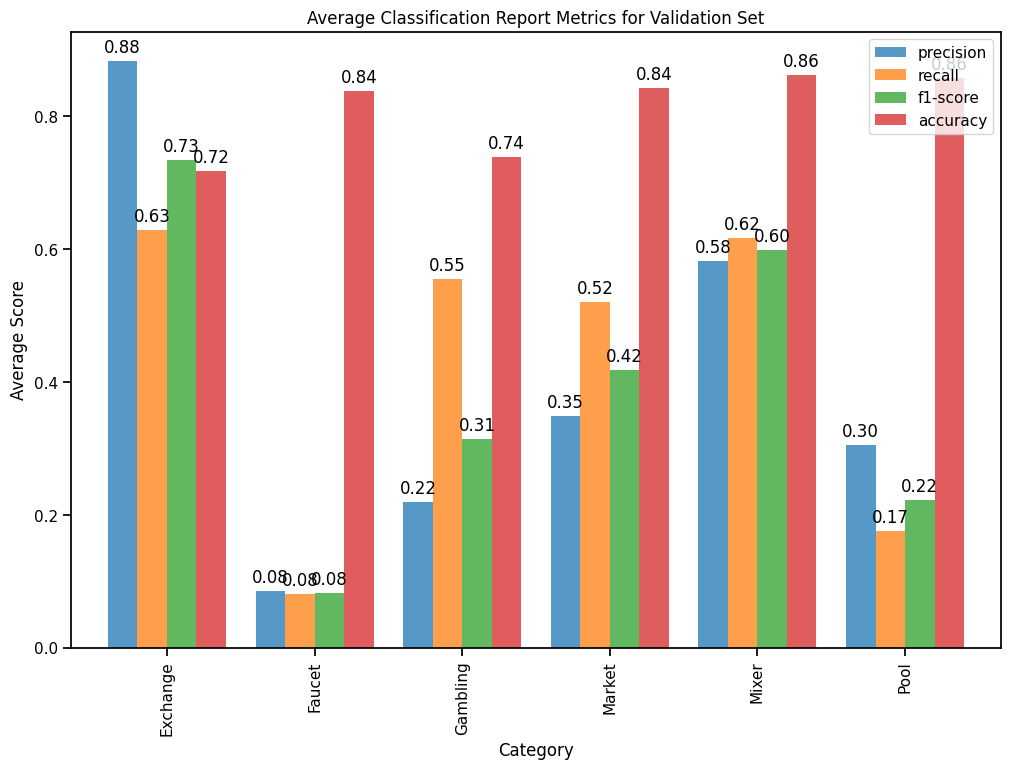}
        \caption{QA Random Forest with Each Class Evaluation}
        \label{fig:Quantum Annealing Random Forest with each class evaluation}
    \end{minipage}
    \hfill
    \begin{minipage}{0.45\textwidth}
        \centering
        \includegraphics[width=\textwidth]{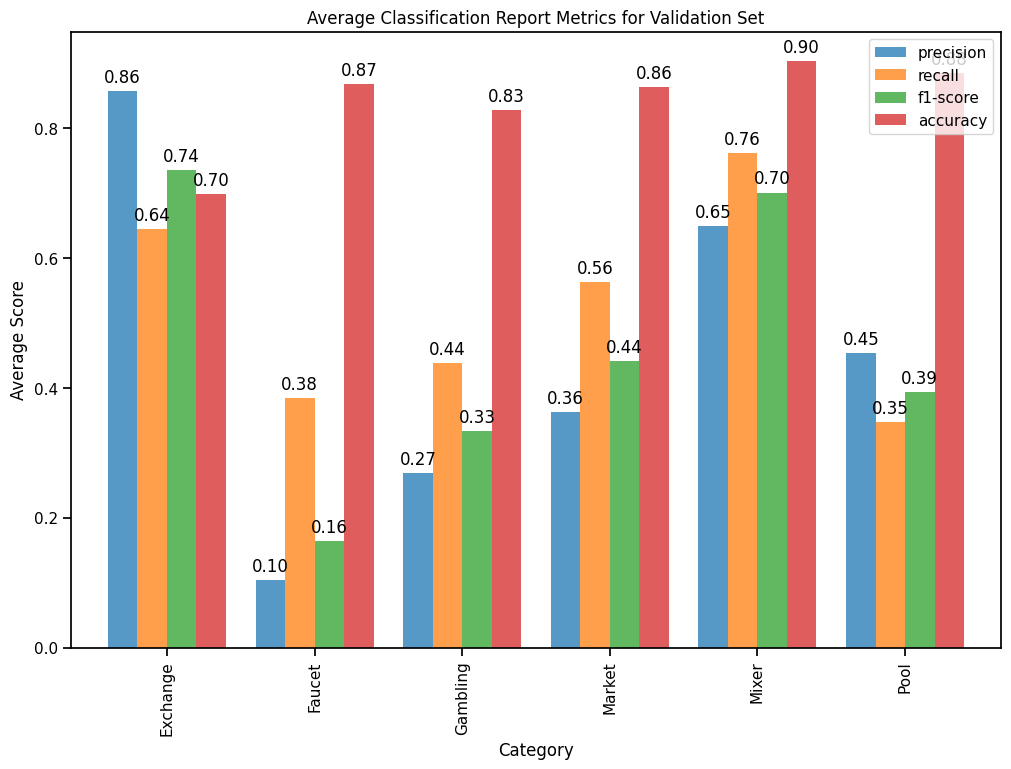}
        \caption{QA BQM Random Forest with Each Class Evaluation}
        \label{fig:Quantum Annealing Binary Quadratic Model Random Forest with each class evaluation}
    \end{minipage}
\end{figure}

The QA Random Forest model, as shown in Figure~\ref{fig:Quantum Annealing Random Forest with each class evaluation}, exhibited more varied results. While it achieved moderate success in the Exchange class with a precision of 0.88\%, it struggled with a much lower recall 0.63\%, yielding an F1-score of 0.73\%. Performance significantly declined in the Faucet class, where both precision 0.08\% and recall 0.08\% were extremely low. In the Gambling class, the Quantum Annealing Random Forest achieved a recall of 0.55\% but at the expense of precision 0.22\%, resulting in an F1-score of 0.31\%. The model also struggled in the Market class, with a precision of 0.35\% and recall of 0.52\%, leading to an F1-score of 0.42\%. Performance for the Mixer class was moderate, with a precision of 0.58\%, recall of 0.62\%, and an F1-score of 0.60\%, and similar issues were seen in the Pool class, where the model achieved a precision of 0.30\% and recall of 0.17\%.

% \begin{figure}[htbp]
% \centerline{\includegraphics[width=15cm]{figures/qarm_testing_recall_without_avg.png}}
% \caption{QA Random Forest with Each Class Evaluation}
% \label{fig:Quantum Annealing Random Forest with each class evaluation}
% \end{figure}

The QA BQM Random Forest model, as demonstrated in Figure~\ref{fig:Quantum Annealing Binary Quadratic Model Random Forest with each class evaluation}, also had mixed results. It showed better performance than the Quantum Annealing Random Forest in the Exchange class, with a precision of 0.86\% and recall of 0.64\%, resulting in an F1-score of 0.74\%. However, for the Faucet class, the model continued to struggle, with precision 0.10\% and recall 0.38\% leading to a low F1-score of 0.16\%. In the Gambling class, the model’s precision 0.27\% and recall 0.44\% yielded an F1-score of 0.33\%. The Market class saw slightly better results, with precision 0.36\% and recall 0.56\%, leading to an F1-score of 0.44\%. In contrast, the Mixer class had a relatively good performance, with precision of 0.65\%, recall of 0.76\%, and an F1-score of 0.70\%. The Pool class showed an improvement in precision 0.45\% but lower recall 0.35\%, yielding an F1-score of 0.39\%.

In conclusion, both Random Forest and SA Random Forest demonstrated better overall classification performance, particularly in achieving a balance between precision and recall for the Exchange and Mixer classes. In contrast, QA Random Forest and QA BQM Random Forest struggled in certain classes, such as Gambling and Faucet, and may require further optimization to improve their precision and recall in those categories.

% \begin{figure}[htbp]
% \centering
% \begin{minipage}{.48\textwidth}  % Adjusted from .5 to .48 to create some space
%   \centering
%   \includegraphics[width=\linewidth]{figures/rm_testing_recall_without_avg.png}
%   \caption{Random Forest with each class evaluation}
%   \label{fig:rf_evaluation}
% \end{minipage}\hfill  % \hfill will push the next minipage to the right, creating space
% \begin{minipage}{.48\textwidth}  % Same adjustment here
%   \centering
%   \includegraphics[width=\linewidth]{figures/qrm_testing_recall_without_avg.png}
%   \caption{Quantum Random Forest with each class evaluation}
%   \label{fig:qrf_evaluation}
% \end{minipage}
% \end{figure}

\subsection{Confusion Matrix}\label{sec:confusion_matrix}
\begin{figure}[htbp]
\centering
\begin{minipage}{.49\textwidth}  % Adjusted for space between images
  \centering
  \includegraphics[width=\linewidth]{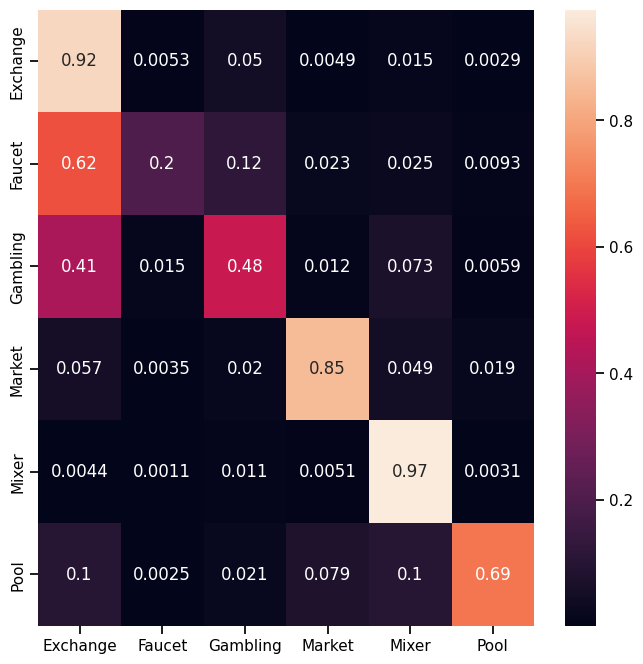}
  \caption{Random Forest Confusion Matrix}
  \label{fig:Random Forest Confusion Matrix}
\end{minipage}\hfill  % Adds horizontal space between the figures
\begin{minipage}{.49\textwidth}
  \centering
  \includegraphics[width=\linewidth]{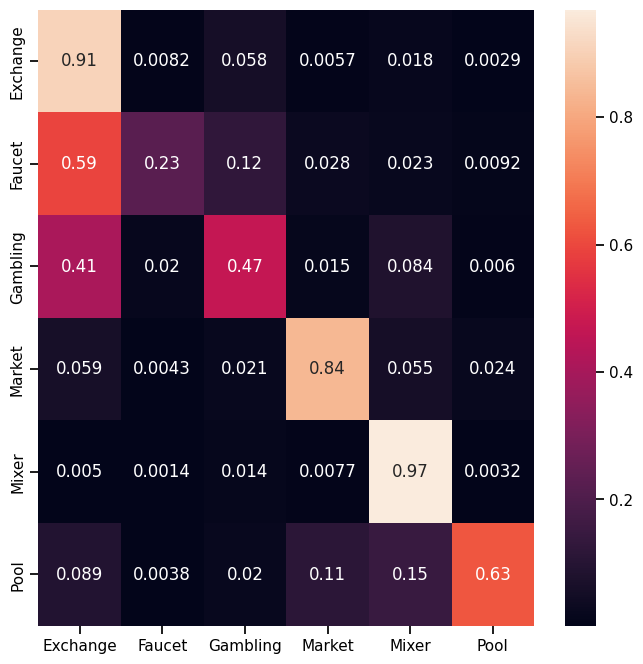}
  \caption{SA Feature Selection Random Forest Confusion Matrix}
  \label{fig:Simulated Annealing Random Forest Confusion Matrix}
\end{minipage}
\end{figure}

\begin{figure}[htbp]
\centering
\begin{minipage}{.49\textwidth}  % Adjusted for space between images
  \centering
  \includegraphics[width=\linewidth]{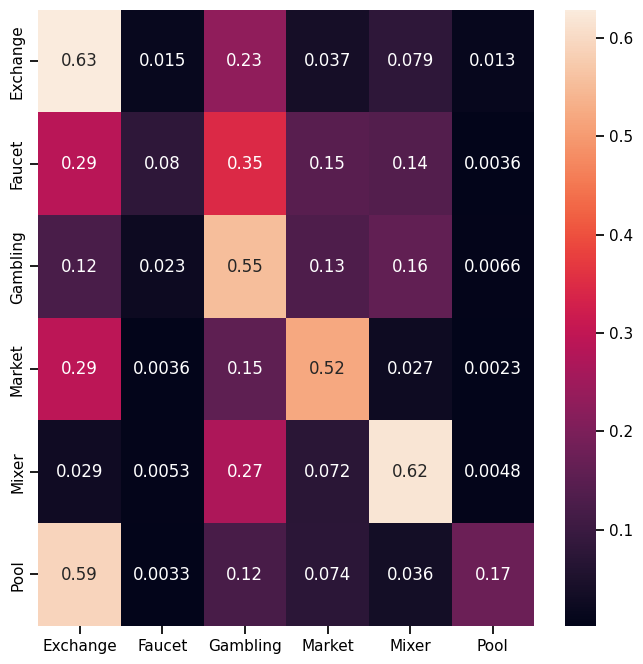}
  \caption{QA Feature Selection Random Forest Confusion Matrix}
  \label{fig:Quantum Annealing Random Forest Confusion Matrix}
\end{minipage}\hfill  % Adds horizontal space between the figures
\begin{minipage}{.49\textwidth}
  \centering
  \includegraphics[width=\linewidth]{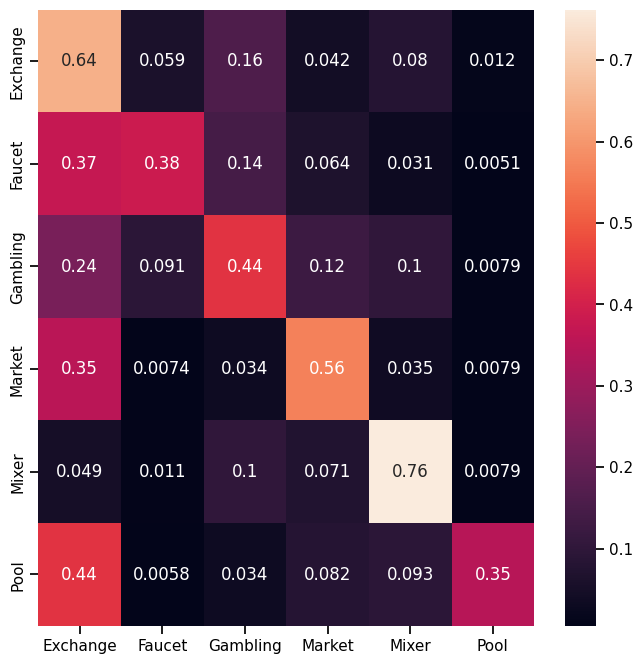}
  \caption{QA BQM Feature Selection Random Forest Confusion Matrix}
  \label{fig:Quantum Annealing BQM Random Forest Confusion Matrix}
\end{minipage}
\end{figure}

We evaluate recall scores for each class within Bitcoin labeled Bitcoin addresses using several supervised machine learning classifiers, alongside SA, QA, and QA BQM quantum algorithms. These classifiers are assessed based on transaction history summaries of labeled addresses. Recall serves as a metric to quantify each model’s effectiveness in accurately identifying positive cases, where a positive case indicates the correct classification of a transaction history as belonging to the mixer class. Figures~\ref{fig:Random Forest Confusion Matrix}, Figure~\ref{fig:Simulated Annealing Random Forest Confusion Matrix},  Figure~\ref{fig:Quantum Annealing Random Forest Confusion Matrix}, Figure~\ref{fig:Quantum Annealing BQM Random Forest Confusion Matrix} illustrate the recall scores for each classifier across distinct classes. Notably, within the mixer class, both the Random Forest and SA Random Forest classifiers achieve the highest recall score of 97\%, indicating their superior effectiveness in identifying mixer transactions. In contrast, QA and QA BQM Random Forest classifiers demonstrate lower recall performances, achieving scores of 0.62\% and 0.76\%, respectively.

% Refer to Appendix~\ref{appx:different_confusion_matrix} for more machine learning classifiers recall socres for each class details. 

% \begin{figure}[htbp]
% \centerline{\includegraphics[width=15cm]{figures/rm_testing.png}}
% \caption{Random Forest Confusion Matrix}
% \label{fig:Random Forest Confusion Matrixn}
% \end{figure}

% \subsection{Confusion Matrix}\label{sec:confusion_matrix}
% \begin{figure}[htbp]
% \centerline{\includegraphics[width=15cm]{figures/qrm_testing.png}}
% \caption{Quantum Random Forest Confusion Matrix}
% \label{fig:Quantum Random Forest Confusion Matrixn}
% \end{figure}
% \clearpage

% \begin{center}
% \begin{table*}[t]
% \caption{Results of Supervised Machine Learning's Recall Scores in Each Class.\label{table:results_of_recall_scores_in_each}}
% \small % Smaller font size
% \centering
% \begin{tabular*}{\textwidth}{@{\extracolsep\fill}lcccccc@{\extracolsep\fill}}
%   \toprule
%   \textbf{Method} & \textbf{Exchange} & \textbf{Faucet} & \textbf{Gaming} & \textbf{Market} & \textbf{Mixer} & \textbf{Pool} \\
%   \midrule
%   Logistic Regression & 0.46 & 0.21 & 0.6 & 0.69 & 0.57 & 0.59 \\
%   AdaBoost      & 0.44 & 0.6 & 0.42 & 0.83 & 0.81 & 0.68 \\
%   Random Forest       & \textbf{0.92} & 0.2 & 0.48 & 0.85 & \textbf{0.97} & 0.69 \\
%   XGBoost             & 0.75 & 0.72 & \textbf{0.68} & \textbf{0.87} & 0.89 & \textbf{0.82} \\
%   LightGBM            & 0.72 & \textbf{0.76} & 0.66 & 0.84 & 0.88 & 0.8 \\
%   Quantum Logistic Regression & 0.36 & 0.4 & 0.38 & 0.55 & 0.8 & 0.55 \\
%   Quantum AdaBoost      & 0.48 & 0.73 & 0.17 & 0.53 & 0.65 & 0.61 \\
%   Quantum Random Forest       & 0.91 & 0.23 & 0.47 & 0.84 & \textbf{0.97} & 0.63 \\
%   Quantum XGBoost             & 0.73 & 0.74 & 0.65 & 0.86 & 0.86 & 0.73 \\
%   Quantum LightGBM            & 0.68 & 0.77 & 0.63 & 0.83 & 0.84 & 0.67 \\
%   \bottomrule
% \end{tabular*}
% \end{table*}
% \end{center}

\subsection{Area under the ROC Curve}\label{sec:area_under_ROC_curve}
% AUC (Area Under the ROC Curve) represents the probability that a classifier will rank a randomly chosen positive instance higher than a randomly chosen negative instance. In this context shown in Table ~\ref{table:evaluation_AUC}, a higher AUC signifies a better ability to distinguish between mixer and non-mixer Bitcoin class. Looking at the AUC score in mixer class, Random Forest, XGBoost, Quantum Random Forest at 0.99 outperform other classifiers and almost equal to exchange, market, mining pool Bitcoin classes. This suggests that these Bitcoin classes module are stable and effect. the full details in Appendix ~\ref{appx:different_auc}.

AUC reflects the probability of a classifier ranking a positive instance higher than a negative one. In Table~\ref{table:evaluation_AUC}, a higher AUC for the mixer class indicates a stronger ability to distinguish between mixer and othres Bitcoin labeled addresses. Notably, Random Forest, XGBoost, and SA Random Forest all achieve outstanding AUC scores of 0.99, rivaling those of exchange, market, and mining pool classes. This suggests these Bitcoin class modules exhibit stability and effectiveness. 
% Refer to Appendix~\ref{appx:different_auc} for more each classes AUC details.

Strikingly, among all classifiers evaluated, SA Random Forest and Random Forest emerged alongside XGBoost as the top performers, exceeding an AUC score of 0.99 for the mixer class. This exceptional performance suggests their remarkable ability to identify a high stable models of genuine mixer transactions. Particularly noteworthy, Random Forest and SA Random Forest demonstrates exceptional performance within the mixer class, achieving the best results among all models with a F1-score of 91\% and an AUC of 0.99.

\begin{table*}[t]
\caption{Evaluation of All Features and Full Features and SA QUBO Feature Selection in Mixer Class of Different Supervised Machine Learning AUC score in Each Class.\label{table:evaluation_AUC}}
\footnotesize % 使用更小的字體
\setcounter{footnote}{0} % 重置註腳編號
\begin{center}
\resizebox{\textwidth}{!}{ % 縮放整個表格以適應頁面
\begin{tabular}{@{\extracolsep{\fill}}p{4cm}p{1.5cm}p{1.2cm}p{1.5cm}p{1.2cm}p{1.2cm}p{1.5cm}@{\extracolsep{\fill}}}
  \toprule
  \textbf{Method} & \textbf{Exchange} & \textbf{Faucet} & \textbf{Gambling} & \textbf{Market} & \textbf{Mixer} & \textbf{Pool} \\
  \midrule
  Logistic Regression & 0.89 & 0.74 & 0.72 & 0.87 & 0.92 & 0.91 \\
  AdaBoost      & 0.89 & 0.67 & 0.37 & 0.96 & 0.74 & 0.80 \\
  Random Forest       & \textbf{0.96} & 0.92 & \textbf{0.90} & \textbf{0.99} & \textbf{0.99} & \textbf{0.98} \\
  XGBoost             & \textbf{0.96} & \textbf{0.93} & 0.89 & \textbf{0.99} & \textbf{0.99} & \textbf{0.98} \\
  LightGBM            & 0.94 & 0.93 & 0.88 & 0.97 & 0.98 & 0.94 \\
  SVM *10\%\footnote{Using only 10\% of the dataset}          & 0.89 & 0.75 & 0.78 & 0.96 & 0.90 & 0.86 \\
  Neural Network *10\%\footnote{Neural Network trained with 10\% of the available data}            & 0.94 & 0.89 & 0.85 & 0.97 & 0.98 & 0.96 \\
  SA Logistic Regression & 0.81 & 0.82 & 0.63 & 0.84 & 0.83 & 0.85 \\
  SA AdaBoost      & 0.85 & 0.61 & 0.37 & 0.91 & 0.69 & 0.79 \\
  SA Random Forest       & 0.95 & 0.91 & 0.88 & \textbf{0.99} & \textbf{0.99} & 0.97 \\
  SA XGBoost             & 0.85 & \textbf{0.93} & 0.88 & \textbf{0.99} & 0.98 & 0.97 \\
  SA LightGBM            & 0.91 & 0.92 & 0.85 & 0.97 & 0.96 & 0.92 \\
  SA SVM *10\%\footnote{SVM trained with 10\% of the dataset using SA}            & 0.72 & 0.71 & 0.77 & 0.79 & 0.71 & 0.68 \\
  SA Neural Network *10\%\footnote{Neural Network trained with 10\% of the data using SA}            & 0.79 & 0.83 & 0.79 & 0.88 & 0.88 & 0.89 \\ 
  QA Logistic Regression & 0.76 & 0.63 & 0.74 & 0.82 & 0.82 & 0.76 \\
  QA AdaBoost      & 0.76 & 0.52 & 0.52 & 0.79 & 0.84 & 0.72 \\
  QA Random Forest       & 0.79 & 0.80 & 0.78 & 0.88 & 0.93 & 0.85 \\
  QA XGBoost             & 0.79 & 0.84 & 0.79 & 0.91 & 0.90 & 0.90 \\
  QA LightGBM            & 0.79 & 0.88 & 0.81 & 0.93 & 0.89 & 0.81 \\
  QA SVM *10\%\footnote{SVM trained with 10\% of the dataset using QA}            & 0.73 & 0.68 & 0.76 & 0.70 & 0.86 & 0.70 \\
  QA Neural Network *10\%\footnote{Neural Network trained with 10\% of the data using QA}            & 0.79 & 0.83 & 0.79 & 0.88 & 0.88 & 0.89 \\  
  QA BQM Logistic Regression & 0.65 & 0.72 & 0.67 & 0.76 & 0.84 & 0.85 \\
  QA BQM AdaBoost      & 0.75 & 0.70 & 0.57 & 0.77 & 0.89 & 0.78 \\
  QA BQM Random Forest       & 0.79 & 0.80 & 0.78 & 0.88 & 0.93 & 0.85 \\
  QA BQM XGBoost             & 0.79 & 0.84 & 0.79 & 0.91 & 0.90 & 0.90 \\
  QA BQM LightGBM            & 0.78 & 0.83 & 0.79 & 0.89 & 0.90 & 0.89 \\
  QA BQM SVM *10\%\footnote{SVM trained with 10\% of the dataset using QA}            & 0.69 & 0.72 & 0.76 & 0.74 & 0.86 & 0.81 \\
  QA BQM Neural Network *10\%\footnote{Neural Network trained with 10\% of the data using BQM}            & 0.78 & 0.85 & 0.78 & 0.89 & 0.88 & 0.89 \\ 
  \bottomrule
\end{tabular}
}
\end{center}
\end{table*}

\subsection{Important Features}\label{sec:evaluation_important_features}

% This section delves into the key features identified by both the Random Forest and Quantum Random Forest models.  Table~\ref{table:top_RM_20_features} showcases the 20 most influential features for the Random Forest classifier, which achieved the best performance. These features are ranked in descending order based on their information gain importance, as described by Louppe et al\cite{louppe2013understanding}.

% Following the successful application of quantum SA's QUBO energy formulation for feature selection, Table~\ref{table:top_QRM_9_features} presents the 9 most important features identified by the Quantum Random Forest classifier. Notably, this model achieves a performance level comparable to the best-performing Random Forest model.

We examine the most important features in different machine learning models, ranked based on their importance of information gain as described by Louppe et al.\cite{louppe2013understanding}. The random forest model shown in Table~\ref{table:top_RM_20_features}, the top 10 most important features of the random forest model are dominated by those related to transaction balances, including USD and BTC, and the transaction volume includes the number spent and received. This suggests that the model heavily relies on financial activity for classification. Features related to transaction patterns and moments like average and standard deviation of balances also appear within the top 20, indicating their influence on the model's predictions.

SA Random Forest model shown in Table~\ref{table:top_QRM_9_features}, the top 9 most important features show some overlap with the Random Forest model, with features related to transaction balances including USD and BTC, spending include USD and BTC appearing again. However, the SA Random Forest model seems to place less emphasis on transaction volume include number spent and received compared to the classical model.
Interestingly, a new feature, lifetime, emerges as important in the SA Random Forest model, suggesting that the model might benefit from understanding the overall duration of user activity.

\begin{table}[t]
\caption{Top 20 Important Features from Random Forest Models.\label{table:top_RM_20_features}}
\centering
\begin{minipage}{0.45\linewidth}
\centering
\textbf{All Features}\\[1ex] % Add a little space after the text
\begin{tabular}{lr}
  \toprule
  \textbf{Feature Name} & \textbf{Feature Type} \\
  \midrule
  $\mu_\mathrm{balance\_usd}$ & Extra Stats \\
  $\mu_\mathrm{balance\_btc}$ & Extra Stats \\
  ${\textit{USD}_\mathrm{received}}$ & Extra Stats \\
  ${\textit{USD}_\mathrm{spent}}$ & Extra Stats \\
  ${\textit{BTC}_\mathrm{spent}}$ & Extra Stats \\
  ${\textit{BTC}_\mathrm{received}}$ & Basic Stats \\
  $\sigma_\mathrm{balance\_btc}$ & Extra Stats \\
  $\sigma_\mathrm{balance\_usd}$ & Extra Stats \\
  $n_\mathrm{spent}$ & Extra Stats \\
  $\mathit{t}_\mathrm{n,output}$ & TX Pattern \\
  \bottomrule
\end{tabular}
\end{minipage}
% \hfill
\begin{minipage}{0.45\linewidth}
\centering
\textbf{The top 11 to 20 features.}\\[1ex]
\begin{tabular}{lr}
  \toprule
  \textbf{Feature Name} & \textbf{Feature Type} \\
  \midrule
  $f_\mathrm{TX}$ & Basic Stats \\
  $\mathit{t}_\mathrm{n,input}$ & TX Pattern \\
  $\mathit{m}_\mathrm{2,received}$ & Moments \\
  $\mathit{m}_\mathrm{2,spent}$ & Moments \\
  $\mathit{m}_\mathrm{2,payback}$ & Moments \\
  $\mathit{m}_\mathrm{2,overall}$ & Moments \\
  $n_\mathrm{received}$ & Extra Stats \\
  $\mathit{m}_\mathrm{1,payback}$ & Extra Stats \\
  $\mathit{m}_\mathrm{1,overall}$ & Extra Stats \\
  $\mathit{m}_\mathrm{1,received}$ & Extra Stats \\
  \bottomrule
\end{tabular}
\end{minipage}
\end{table}

\begin{table}[t]
\caption{Top Important Features from SA Random Forest Models.\label{table:top_QRM_9_features}}
\centering
\textbf{SA Model Features}\\[1ex]
\begin{tabular}{lr}
  \toprule
  \textbf{Feature Name} & \textbf{Feature Type} \\
  \midrule
  ${\textit{BTC}_\mathrm{spent}}$ & Extra Stats \\
  $\mu_\mathrm{balance\_usd}$ & Extra Stats \\
  ${\textit{USD}_\mathrm{spent}}$ & Extra Stats \\
  ${\textit{USD}_\mathrm{received}}$ & Extra Stats \\
  $\sigma_\mathrm{balance\_btc}$ & Extra Stats \\
  $\textit{lifetime}$ & Extra Stats \\
  $\mathit{m}_\mathrm{1,interval}$ & Moments \\
  $n_\mathrm{TX}$ & Extra Stats \\
  $n_\mathrm{payback}$ & Extra Stats \\
  \bottomrule
\end{tabular}
\end{table}

QA Random Forest model shown in Table~\ref{table:combined_features}, the top 6 most important features highlight the significance of received transactions and various statistical moments. In particular, the feature related to the USD received reappears, indicating its importance across models. Basic statistics such as transaction frequency also emerge, alongside higher moments that may provide deeper insights into the distribution of transaction amounts. Additionally, transaction patterns are emphasized through features that capture output behavior and overall transaction trends, suggesting that a holistic view of user activity is essential in this model.

\begin{table}[t]
\caption{Top Important Features from QA Random Forest Models.\label{table:combined_features}}
\centering
\begin{minipage}{0.45\linewidth}
\centering
\textbf{QA Model Features}\\[1ex]
\begin{tabular}{lr}
  \toprule
  \textbf{Feature Name} & \textbf{Feature Type} \\
  \midrule
  ${\textit{USD}_\mathrm{received}}$ & Extra Stats \\
  $f_\mathrm{TX}$ & Basic Stats \\
  $\mathit{m}_\mathrm{2,interval}$ & Moments \\
  $\mathit{m}_\mathrm{4,received}$ & Moments \\
  $\mathit{t}_\mathrm{n,output}$ & TX Pattern \\
  $\mathit{t}_\mathrm{overall}$ & TX Pattern \\
  \bottomrule
\end{tabular}
\end{minipage}
\hfill
\begin{minipage}{0.45\linewidth}
\centering
\textbf{QA BQM Model Features.}\\[1ex]
\begin{tabular}{lr}
  \toprule
  \textbf{Feature Name} & \textbf{Feature Type} \\
  \midrule
  $\mu_\mathrm{balance\_btc}$ & Extra Stats \\
  $f_\mathrm{TX}$ & Basic Stats \\
  $\textit{lifetime}$ & Extra Stats \\
  $\mathit{m}_\mathrm{3,payback}$ & Moments \\
  $\mathit{t}_\mathrm{overall}$ & TX Pattern \\
  \bottomrule
\end{tabular}
\end{minipage}
\end{table}

% \begin{table}[t]
% \caption{The top 6 Important Features on QA Random Forest Model.\label{table:top_QARM_6_features}}
% \centering
% \textbf{QA 6 features.}\\[1ex]
% \begin{tabular}{lr}
%   \toprule
%   \textbf{Feature Name} & \textbf{Feature Type} \\
%   \midrule
%   ${\textit{USD}_\mathrm{received}}$ & Extra Stats \\
%   $f_\mathrm{TX}$ & Basic Stats \\
%   $\mathit{m}_\mathrm{2,interval}$ & Moments \\
%   $\mathit{m}_\mathrm{4,received}$ & Moments \\
%   $\mathit{t}_\mathrm{n,output}$ & TX Pattern \\
%   $\mathit{t}_\mathrm{overall}$ & TX Pattern \\
%   \bottomrule
% \end{tabular}
% \end{table}

In the QA BQM Random Forest model, the top 5 important features reveal a continued emphasis on balance statistics and lifetime as significant factors. The average balance in BTC demonstrates its relevance, along with the transaction frequency as a basic statistic. The inclusion of the lifetime feature further aligns with the trend observed in other models, indicating its potential impact on predicting user behavior. Moments related to payback and overall transaction patterns are also included, reinforcing the model's focus on understanding transactional dynamics over time.

% \begin{table}[t]
% \caption{The top 5 Important Features on QA BQM Random Forest Model.\label{table:top_QABQMRM_5_features}}
% \centering
% \textbf{QA BQM 5 features.}\\[1ex]
% \begin{tabular}{lr}
%   \toprule
%   \textbf{Feature Name} & \textbf{Feature Type} \\
%   \midrule
%   $\mu_\mathrm{balance\_btc}$ & Extra Stats \\
%   $f_\mathrm{TX}$ & Basic Stats \\
%   $\textit{lifetime}$ & Extra Stats \\
%   $\mathit{m}_\mathrm{3,payback}$ & Moments \\
%   $\mathit{t}_\mathrm{overall}$ & TX Pattern \\
%   \bottomrule
% \end{tabular}
% \end{table}

\section{Conclusion}

In this work, we propose a novel framework for identifying Bitcoin mixer addresses by integrating QIFS with transaction history summaries and supervised machine learning classifiers. Our approach utilizes both existing statistical measures and newly introduced Transaction Pattern features, which significantly contribute to improved classification accuracy, particularly with the Random Forest model.

The transaction history summaries encompass essential information about Bitcoin addresses, including basic statistics, extended statistics, moments, and transaction patterns. The basic statistical features build on previous work by Toyoda et al. \cite{toyoda2018multi}, while the extended statistics and moments were inspired by Lin et al. \cite{lin2019evaluation}. The Transaction Patterns features, informed by the MIOT model, were crucial in capturing mixer address activity.

Our experimental evaluation employed a range of classification algorithms, including Logistic Regression, AdaBoost-SAMME, Random Forest, XGBoost, LightGBM, SVM, and Neural Network, using the Scikit-learn Python library. To optimize the classifiers, we applied 10-fold cross-validation. The experiment results demonstrated that Random Forest and SA Random Forest were the best-performing models for the mixer class, achieving F1 scores of 92% and 91%, respectively. Notably, the SA Random Forest model exhibited a significant reduction in training time, offering a 30.3% speed improvement compared to the traditional Random Forest while maintaining comparable accuracy.

We also evaluated the precision, recall, and F1 scores for six different address categories: Exchange, Faucet, Gambling, Market, Mixer, and Pool. Both Random Forest and SA Random Forest achieved high performance in the Mixer class, while QA and QA BQM-based classifiers showed lower effectiveness, indicating room for improvement.

Overall, the combination of QIFS, particularly with SA and Random Forest, demonstrates substantial potential for enhancing the efficiency and accuracy of Bitcoin mixer identification. This approach not only achieves competitive classification metrics but also reduces computational overhead, showcasing the viability of quantum-inspired feature selection in financial transaction analysis tasks.

\subsection{Future Work}

The current classification methodology is affected by data imbalance, as highlighted in Table \ref{table:dataset_details}. This imbalance often leads to increased false negatives and reduced sensitivity for minority classes, ultimately degrading the model's overall reliability and robustness. To mitigate this limitation, we plan to integrate the Synthetic Minority Oversampling Technique (SMOTE) \cite{chawla2002smote} prior to feature summarization. This technique is anticipated to enhance the recall for underrepresented classes, resulting in a more balanced and robust classification model. Additionally, incorporating cost-sensitive learning during the training phase of Random Forest and other decision tree-based algorithms could further improve model performance, as indicated by Seliya et al. \cite{Seliya2011The}.

We also propose conducting an ablation study to evaluate the contribution of each feature set to the overall model performance. Understanding the impact of individual feature sets will inform future improvements by identifying which features are most crucial for classification accuracy and which are redundant, allowing for a more efficient and streamlined model. This study will involve systematically removing different feature groups—such as Transaction Patterns—and assessing the resultant impact on classification accuracy. By quantifying the influence of individual feature sets and examining the effectiveness of quantum-inspired SA in feature selection, we aim to gain a deeper understanding of the factors driving performance improvements.

Moreover, the QIFS methodology demonstrates substantial potential beyond its current scope, which is currently focused on Bitcoin mixer address classification. By optimizing feature selection, QIFS can significantly reduce computational overhead for classical machine learning models, thereby facilitating efficient processing across a variety of domains where feature selection plays a critical role. This positions QIFS as a viable approach for broader applications in fields such as cybersecurity, healthcare, and predictive maintenance, where effective feature selection is pivotal for improving model efficiency and overall performance. For instance, in cybersecurity, QIFS can be used to identify key features for detecting network intrusions, enhancing the precision and speed of intrusion detection systems. In healthcare, QIFS could optimize the selection of biomarkers from high-dimensional genomic data, thus improving the accuracy of diagnostic models. In predictive maintenance, QIFS can help identify critical indicators of machinery health, allowing for proactive interventions and minimizing downtime.

\bmhead{Acknowledgements}
We acknowledge support from the National Science and Technology Council, Taiwan, under Grants NSTC
112-2119-M-033-001, by the research project Applications of quantum computing in optimization and finances.

\bibliography{sn-bibliography}% common bib file

\end{document}